\documentclass[
    ,final            
  ]
  {aipproc}

\layoutstyle{6x9}

\begin{document}

\title{Electron Spin Precession at CEBAF}


\author{D. W. Higinbotham}{
  address={Jefferson Lab, 12000 Jefferson Ave., Newport News, VA 23606}
}

\begin{abstract}

The nuclear physics experiments at the Thomas Jefferson National Accelerator Facility often require longitudinally polarized
electrons to be simultaneously delivered to three experimental halls.  The degree of longitudinal polarization to each hall varies as function of the accelerator settings, making it challenging in certain situations  to deliver a high degree of longitudinal polarization to all the halls simultaneously.   
Normally, the degree of longitudinal polarization the halls receive is optimized by changing the initial spin direction  at the beginning of the machine with a Wien filter.
Herein, it is shown that it is possible  to  further improve the degree of longitudinal polarization for multiple experimental halls by redistributing the energy gain of the CEBAF linacs while
keeping the total energy gain fixed.
 \end{abstract}

\maketitle


\section{Introduction}

Longitudinally polarized electron beams are produced at Jefferson Lab with a strained superlattice GaAs wafer and a DC photogun~\cite{HernandezGarcia:2008zz,Sinclair:2007ez}.  These polarized electrons are required for a large fraction of the Jefferson Lab nuclear physics experiments; but the spin precession from the injector to the experimental halls is complicated by the fact that the accelerator runs with different  total energies, different number of passes around the machine for different halls, and even with different bend angles into the various halls.   In the past, once a beam energy was established, the only parameter that could be changed to manipulate the spin was the Wien angle at the beginning of the machine~\cite{Leemann:2001dg,0034-4885-68-9-R01}.  If only one hall required polarization, by adjusting the Wien angle, that hall could receive the full injected longitudinal polarization.  The challenge is to provide the highest possible longitudinal  polarization simultaneously to multiple halls.

\section{Spin Precession}
\begin{figure}[hbt]
\includegraphics[width=\textwidth]{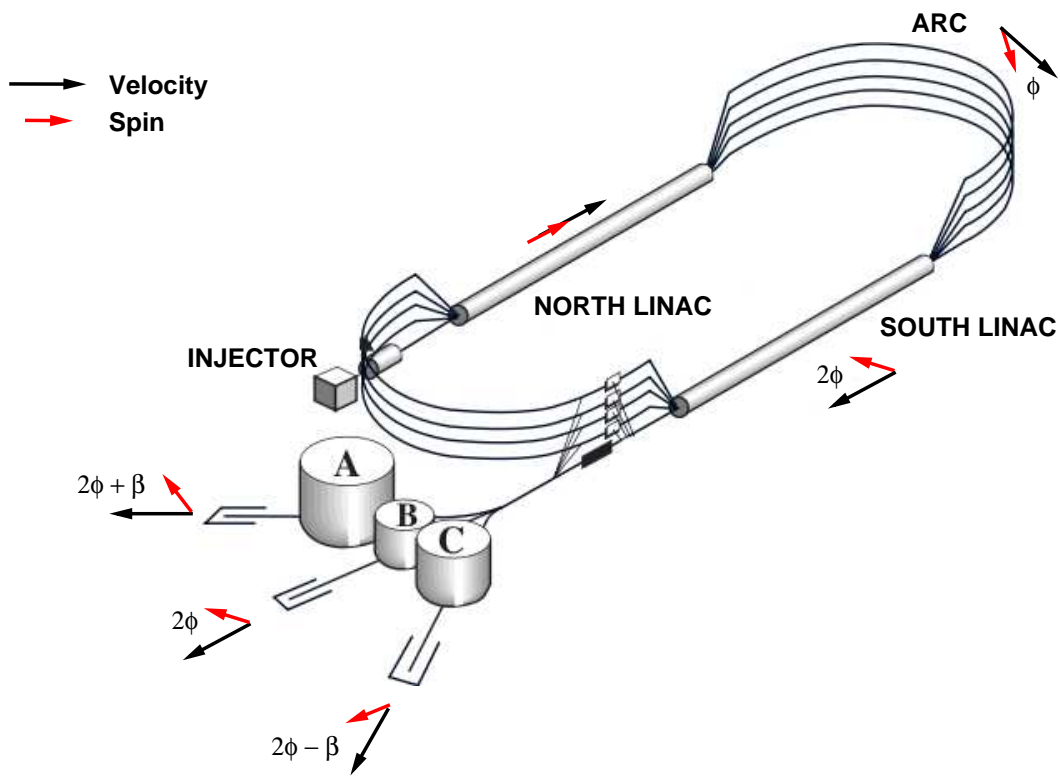}
\caption{Shown is a simple example of spin precession in CEBAF.  In this example first pass beam is being sent to all three halls.  The velocity and spin directions start out parallel at the injector and continue parallel through the north linac.  As the beam passes through the arc, spin precession causes the two vectors to start to point in different directions.  By the time the beam has arrived at the three different end stations, the angle between the velocity and spin has taken the values of $2\phi + \beta$, $2\phi$, and $2\phi - \beta$, for Hall A, B, and C, respectively.}
\label{fig:spin-fig1}
\end{figure}

In general, the spin precession, $\Delta\phi$, of an electron of fixed energy as it  bends through a series of dipole magnets is given by:
\begin{equation}
\Delta\phi = \frac{\mathrm{E}}{440.65 \mathrm{MeV}} \times \Delta\theta
\end{equation}
where E is the energy of the electron and $\Delta\theta$ is the bend angle~\cite{Montague:1983yi}.
A simple example of CEBAF spin precession for first pass beam being hypothetically delivered to all three halls is shown in 
Fig.~\ref{fig:spin-fig1}.  In this example, the spin direction starts at the injector along the velocity direction ends in the three end stations with angles of $2 \phi + \beta$, $2 \phi$, and $2 \phi - \beta$ with respect to the velocity direction. 
For even this simple calculation, two different energies need to be taken into account as the electrons increase in energy for each
pass through a linac.    In the example, the spin  precesses by $\phi$ to the midpoint of the arc and by then another $\phi$ as it completes its bend around the arc all with an energy $E_{north~linac}$; but after passing through the
south linac the precession needs to be calculated of an energy of $E_{north~linac} + E_{south~linac}$  and 
a bend angle of  37$^{\circ}$, 0$^{\circ}$, -37$^{\circ}$ bend for Hall A, B, or C, respectively.   
Since Jefferson Lab is a race track accelerator, the energies continue to increase for higher passes.  Nevertheless,  a simple series of precession calculations are sufficient to calculate the total precession to any given Hall and at given pass.   In fact, this relation between the beam energy and the amount of precession has been used to help determine the beam energy at Jefferson Lab~\cite{Grames:2004mk}.

\section{Linac Energy Imbalancing}

\begin{figure}[htb]
\includegraphics[width=\textwidth]{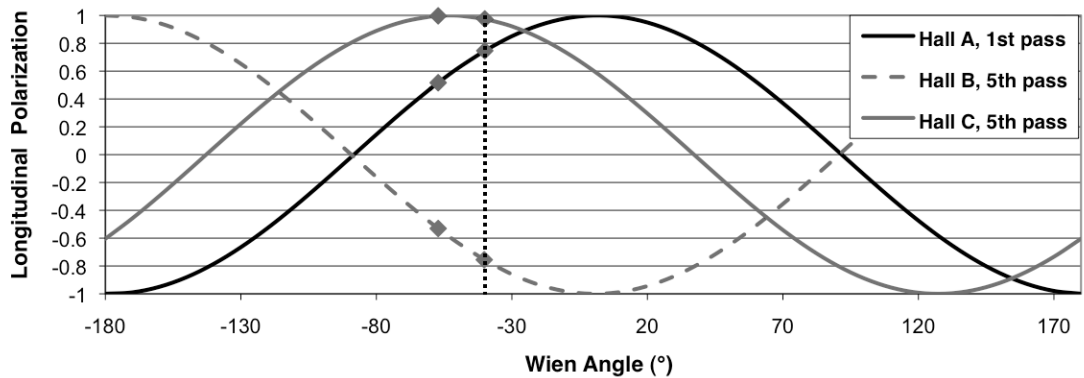}
\caption{Shown is the degree of longitudinal polarization to the end stations for CEBAF setup with the injector set to 63.5 MeV and each of the linacs set to a gain of 565 MeV per pass.  The points represent normalized Moller measurements done by the various Halls at different Wien settings.  During this run period, in the summer of 2008, Hall C was the only experiment running which required the highest polarization.  The dotted line indicates the Wien angle setting of -40$^{\circ}$ for this run period.}
\label{fig:spin-balanced}
\end{figure}

In the summer of 2008, the CEBAF accelerator was running with only Hall C requiring high longitudinal polarization.  The results of Moller measurements done in the halls during this time along with curves showing the expected  degree of longitudinal polarization as a function of Wien angle is shown in Fig.~\ref{fig:spin-balanced}.  
For spin experiments that use only beam polarization, like the one that was running in Hall C, the figure of merit of the experiment is proportional to P$^{2}$, the polarization square.

\begin{figure}[htb]
\includegraphics[width=\textwidth]{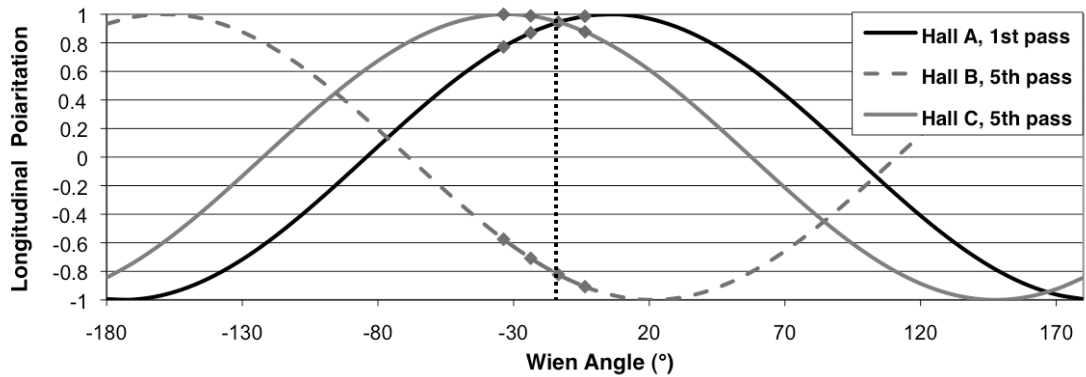}
\caption{By reducing the north linac by 10 MeV and increasing the south linac by 10 MeV, we were able to deliver high polarization to both Hall's A $\&$ C for the second half of the summer 2008 run period while not changing the beam energy Hall B was receiving.  The points represent normalized Moller measurements done by the various Halls at different Wien settings.  The dotted line indicates the Wien angle setting of -13.5$^{\circ}$ that was used after the linacs were imbalanced. }
\label{fig:spin-unbalanced}
\end{figure}

During the second half of this same run period, both Hall A $\&$ C required high polarization while Hall B required that the beam energy stay fixed.   In the past, the Wien angle was the only parameter to change; but this would have only provided a  P$^2$ of only 0.8 to Halls A $\&$ C and thus effectively caused a 10$\%$ reduction in beam time compared to the P$^2$ of 0.9 they were expecting.
By running with imbalanced energy gains in the north and south linacs, the accelerator was able to precess the spin in such a may as to provide a P$^2$ of 0.9 to both Halls A $\&$ C while still maintaining the same total energy.  The effect on the spin precession of the imbalancing is shown in Fig.~\ref{fig:spin-unbalanced} and summarized in Table~\ref{tab:spin-table}.

\begin{table}[ht]
\begin{tabular}{ccccc}
\hline
  & \tablehead{1}{r}{b}{Injector}
  & \tablehead{1}{r}{b}{North Linac}
  & \tablehead{1}{r}{b}{South Linac}
  & \tablehead{1}{r}{b}{P$^2$}   \\
 \hline

Balanced       &  63.5  MeV  & 565 MeV     & 565 MeV   &  0.8 \\
Unbalanced  &  63.5 MeV   & 555 MeV     & 575 MeV   &  0.9 \\ \hline
\end{tabular}
\caption{Shown are the conditions before and after imbalancing the CEBAF linacs.  With Halls A and C both doing polarization transfer experiments, being able to keep both Halls at a polarization squared, P$^2$, of 0.9 prevented an effective 10\% reduction in beam time.}

\label{tab:spin-table}

\end{table}

Now that this method of optimizing the spin precession for multiple halls 
by imbalancing the energy gain of the accelerator's north and south linacs has been demonstrated, Jefferson Lab's accelerator
division is determining its potential for future experiments.
With two degrees of freedom, the Wien angle and linac imbalance, the physics experiments should receive,
for the same beam energy, higher longitudinal polarization then was previously achievable.

%

\begin{theacknowledgments}
Special thanks to my summer student Marie-Isabelle Holdrinet for her work on this topic and to the Jefferson Science Associates initiatives fund that make her stay at Jefferson Lab possible.  And thanks Arne Freyberger and the Jefferson Lab accelerator scientists for their collaboration and for turning the idea of running the machine imbalanced into a reality.
\end{theacknowledgments}



\bibliographystyle{aipproc}   

\bibliography{douglas-higinbotham-spin2008}

\end{document}